# The Li Intercalation Potential of LiMPO$_4$ and LiMSiO$_4$ olivines with M = Fe, Mn, Co, Ni

Fei Zhou,
*Department of Physics, Massachusetts Institute of Technology, Cambridge, MA 02139, USA*

Matteo Cococcioni, Kisuk Kang, Gerbrand Ceder*
*Department of Materials Science and Engineering, Massachusetts Institute of Technology, Cambridge, MA 02139, USA*

August 23, 2004

**Abstract**

The Li intercalation potential of LiMPO$_4$ and LiMSiO$_4$ compounds with M = Fe, Mn, Co, and Ni is computed with the GGA+*U* method. It is found that this approach is considerably more accurate than standard LDA or GGA methods. The calculated potentials for LiFePO$_4$, LiMnPO$_4$ and LiCoOPO$_4$ agree to within 0.1 V with experimental results. The LiNiPO$_4$ potential is predicted to be above 5 V. The potentials of the silicate materials are all found to be rather high, but LiFeSiO$_4$ and LiCoSiO$_4$ have negligible volume change upon Li extraction.

## Introduction

First principles computations have shown to be relevant for predicting many of the properties of Li-insertion materials used as electrodes in rechargeable Li batteries [1-9]. One of the critical properties for a Li intercalation material is the potential at which Li can be removed and inserted. While a high potential increases the energy density of the material, if this potential is too high, Li can not be practically removed, and side reactions such as electrolyte breakdown, can occur in the cell. A low potential can also lead to moisture sensitivity of the electrode material. Hence, knowledge of the thermodynamic potential is one crucial aspect when determining whether new materials can be used as cathode materials in rechargeable lithium batteries. In this communication, we predict the potential of LiNiPO$_4$, LiMnSiO$_4$, LiFeSiO$_4$, LiCoSiO$_4$ and LiNiSiO$_4$ using the highly accurate GGA+*U* method.

The electrochemical activity of LiFePO$_4$ [10] and Li$_3$V$_2$(PO$_4$)$_3$ [11-14] has spawned considerable interest in materials with poly-anion groups, as they may be considerably more stable than close-packed oxides at the end of charging. Currently, LiFePO$_4$ and Li$_3$V$_2$(PO$_4$)$_3$ are of commercial interest, though the rather low potential of LiFePO$_4$ gives it too low an energy density to compete with layered oxides in applications where volumetric or gravimetric energy density is most important. In this communication we use ab-initio computations to predict the potential of other phosphates and silicates in the olivine structure. While ab-initio methods have previously been used to predict the potential of insertion electrodes, we use a more accurate quantum mechanical approach in this work, allowing us to predict the insertion potentials within 0.1-0.2V. We find that of the compounds investigated, LiFePO$_4$ actually has the lowest potential. Among the phosphate compounds, the potential of LiNiPO$_4$ is found to be too high to be of practical interest. LiMnSiO$_4$, LiFeSiO$_4$, LiCoSiO$_4$ and LiNiSiO$_4$ also have high potentials, with only the Mn and Fe material just below 5V.

*Corresponding Author: gceder@mit.edu
+1-617-253-1581 (phone) 617-258-6534 (fax)

## Methods

It has previously been show that the average potential for Li extraction from a material is given by:

$$V = - \frac{G[\text{Li}_{x_2}Host] - G[\text{Li}_{x_1}Host] - (x_2 - x_1)G[\text{Li}]}{x_2 - x_1} \quad (1)$$

where G is the Gibbs free energy of the compound. Typically, the equilibrium Li potential in insertion materials $\text{Li}_x Host$ varies with composition $x$. The above equation gives the average of the equilibrium potential between the compositions $x_2$ and $x_1$. While it is possible to study the detailed variation of the potential with composition [5], the average potential usually contains enough information to decide whether a material is of practical use in rechargeable Li batteries. Typically, the free energies can be replaced by the ground state energies with very little error [1]. Hence, to determine an average intercalation potential the energy of only three compounds needs to be determined: $\text{Li}_{x1}Host$, $\text{Li}_{x2}Host$ and metallic Li.

In almost all previous ab-initio studies these energies have been determined with standard Density Functional Theory (DFT) in either the Local Density Approximation (LDA) or Generalized Gradient Approximation (GGA). It is found that typically, these methods underestimate the potential, with errors ranging from 0.25 V to more than 1 V [15]. There is some recent evidence that cancellation of the self-interaction in methods such as LDA+$U$ corrects this error, and can lead to much better voltage predictions [16, 17, 8]. This seems to be particularly important in phosphate materials, where due to a strong localization of the metal $d$-orbitals the LDA and GGA errors are rather large.

The LDA+$U$ method (or GGA+$U$ when GGA is used, or DFT+$U$ in general) was invented to more accurately treat strongly correlated materials, such as transition metal or rare earth compounds [18-20]. The concept of DFT+$U$ is to treat the localized $d$-states by explicit many-body theory, while retaining the regular DFT Hamiltonian (LDA or GGA) for the other more delocalized states. The $d$-part of the Hamiltonian is treated with standard Hubbard models. The effect of this hybrid Hamiltonian is to modify the electron potential within the transition metal or rare earth atom radii from the LDA/GGA by an amount proportional to the $U$ parameter in this model. Compared to GGA, there is little additional computational cost in the GGA+$U$ method, making it an efficient way to study transition metal compounds. The Hubbard model for the $d$-states penalizes partial occupancy of $d$-orbitals, which has been found to be critical for the prediction of phase stability in LiFePO$_4$ [16]. More details on the implementation of the DFT+$U$ method can be found in Refs [16, 17].

For each M = Fe, Mn, Co and Ni the energy of the LiMPO$_4$, MPO$_4$, LiMSiO$_4$ and MSiO$_4$ structures have been computed. All structures were fully relaxed (cell shapes as well as internal coordinates) so as to get the structural parameters that give the lowest energy. We calculated the total energies with GGA (PBE) [21] and GGA+$U$, with the projector-augmented wave (PAW) method [22] as implemented in the Vienna Ab-initio Simulation Package (VASP) [23, 24]. The use of GGA over LDA has previously been shown to be essential for correctly reproducing magnetic interactions and possible Jahn-Teller distortions [3]. We chose an energy cut-off of 500 eV and appropriate $k$-point mesh so that the total ground state energy is converged to within 3meV per formula unit.   Magnetic configuration and ordering is important for determining properties of transition metal compounds.   Each ion was initialized in different magnetic configurations (non-spin polarized, low spin, high spin) in the electronic convergence iterations to get the lowest energy electronic structure configuration.   We have found that initialization in the high spin configuration was sufficient in most of the materials studied.   The ordering of the magnetic spin on the ions, i.e. ferromagnetic (FM), antiferromagnetic (AFM) or more complicated orderings, can results in difference in the total energy of the order 10-60 meV per formula unit. In the phosphates, the spin ordering was initialized as AFM, based on the neutron-diffraction results on several LiMPO4 compounds [25-27]. In the silicates FM ordering was used due to lack of experimental information.   The values of $U$ were determined through a recently developed linear response method and fully consistent with the definition of the LDA+$U$ Hamiltonian [28], making our approach for the potential calculations fully ab-initio.   For each system, $U$ was calculated for the oxidized and reduced ion, and the voltage was taken as the average obtained with both $U$ values.

*Corresponding Author: gceder@mit.edu
+1-617-253-1581 (phone) 617-258-6534 (fax)

## Results

Table 1 shows the values of $U$ obtained for each olivine phosphate compound, the lattice parameters of the fully relaxed configurations, the magnetic moments per transition metal ion, and the average intercalation potential.   For comparison, standard GGA results as well as more accurate GGA+$U$ results are shown.   Experimental values are shown where available.   Results for the silicates are shown in Table 2.   The $U$ values which are a measure of the electron-electron interaction on the transition-metal site show a general increase with increased valence state.   This may reflect the lack of screening as the number of $d$-electrons is reduced.     Only for $Mn^{3+}/Mn^{4+}$ and $Ni^{3+}/Ni^{4+}$ in the silicate systems is the trend opposite.

The potentials are plotted in Figure 1.   As was found in previous systems [1], the Li extraction potential generally increases as one goes to the right in the 3$d$-metal series.   The minimum in potential for LiFePO$_4$ is a notable exception, and is related to the particular electron structure of $Fe^{3+}$ (see below).   The excellent agreement between the calculated and experimental voltage for LiFePO$_4$, LiMnPO$_4$ and LiCoPO$_4$ is a validation of our approach.   The GGA+$U$ significantly improves upon the pure GGA approach ($U = 0$), confirming earlier work [16, 17, 8].   Whereas in GGA errors of 1 V are not uncommon, the largest error for the three systems for which GGA+$U$ can be compared to experiments is less than 0.1 V.   We believe that this improvement is due to the cancellation of the self-interaction on the $d$-orbitals in GGA+$U$.   In standard LDA/GGA electron self-interaction places the occupied $d$-levels too high, and therefore the oxidation potential too low, which leads to a systematic underestimation of the Li extraction potential [1].   For LiNiPO$_4$ no experimental result is known.   We predict a Li intercalation potential of 5.1 for LiNiPO$_4$ which is close to, or above the maximum potential that can be tolerated by Li electrolytes, explaining why delithiation of LiNiPO$_4$ may not have been observed.

The active redox couple in LiMSiO$_4$ materials ($M^{3+}/M^{4+}$) is one valence state higher than for the phosphate.   Hence, on the basis of the higher valence redox couple, one would expect the potentials in the silicates to be higher than those in the phosphates.   In addition, the anion group can also affect the potential through its hybridization with the transition metal [1, 29, 30].   Our results confirm that the potential of the silicate olivines (Fig. 1) is indeed substantially higher than for the corresponding phosphate olivines.     The only potentials predicted to be below 5V are those of LiMnSiO$_4$ (4.8V) and LiFeSiO$_4$ (4.9 V). The Li extraction for LiCoSiO$_4$ and LiNiSiO$_4$ is above 5 V and likely too high to be of any practical value. The potential differences between the phosphates and silicates are not constant, and reflect the electronic structure of the transition metal ions. For Co and Ni, the potential of the silicate and phosphate is almost the same, while for Mn and Fe it is significantly different. The large increase in going from LiFePO$_4$ to LiFeSiO$_4$ is due to the stable shell structure of Fe$^{3+}$. With five $d$-electrons in high spin state, Fe$^{3+}$ forms a half-closed shell. Oxidation of Fe$^{2+}$ (in LiFePO$_4$) forms this closed shell, and hence the potential is low, while oxidation of Fe$^{3+}$ (in LiFeSiO$_4$) to Fe$^{4+}$ destroys this stable electronic configuration. As a result, the Li extraction potential in LiFeSiO$_4$ is 1.5 V higher than in the phosphate. This difference is much less for the Mn compounds. For LiMnPO$_4$ we calculate a potential of 4.04 V, in good agreement with the potential measured in Li(Fe,Mn)PO$_4$ mixtures [31] and in LiMnPO$_4$ [32]. The calculated potential in LiMnSiO$_4$ (4.87 V) is higher by only 0.8 V. Li extraction from LiMnPO$_4$ destroys the formation of the closed shell in Mn$^{2+}$ to make Mn$^{3+}$. In the silicate, LiMnSiO$_4$, no closed shell is destroyed. The difference between the phosphate and silicate potentials decreases for the later transition metals Co and Ni. This could reflect an increased participation of the anion group at these high potentials [1, 33]. At high levels of oxidation, rehybridization shifts occur in the transition-metal ligand bond in order to diminish the effect of changing the valence on the transition metal site. Such shifts have been previously observed with computational methods in LiCoO$_2$ [1, 34] and NaCoO$_2$ [35], and experimentally in LiCoO$_2$ [36].

*Corresponding Author: gceder@mit.edu
+1-617-253-1581 (phone) 617-258-6534 (fax)

The calculated GGA+$U$ volumes for the lithiated and delithiated phosphates and silicates are given in Table 1 and 2 and plotted in Figure 2. Comparison of the calculated volume changes upon delithiation with the experimental values for LiFePO$_4$, LiMnPO$_4$ and LiCoPO$_4$ indicate that GGA+$U$ is in general better at predicting these volume differences. For LiCoPO$_4$ the prediction of volume change with GGA+$U$ ($\Delta V = 5$ Å$^3$/formula unit) is in considerably better agreement with experiment ($\Delta V = 4$ Å$^3$/formula unit) than the prediction with standard GGA ($\Delta V = 11$ Å$^3$/formula unit). This particularly large difference between GGA and GGA+$U$ is due to the different spin configurations obtained in both methods. GGA+$U$ more correctly favors high spin which leads to a larger volume for CoPO$_4$. The spin state is also at the origin of the large volume differences between GGA and GGA+$U$ for LiCoSiO$_4$, CoSiO$_4$ and NiSiO$_4$. The volume changes with delithiation are slightly smaller in the silicates than in the phosphates. In particular, LiFeSiO$_4$ and LiCoSiO$_4$ have almost zero volume change as Li is removed.

**Conclusion**

The GGA+$U$ method predicts Li voltages with remarkably good accuracy in phosphate olivines, in contrast to previously standard ab-initio approaches. As $U$, the electron-electron interaction on the $d$-orbitals can be determined from ab-initio linear response theory [28], the GGA+$U$ approach is still fully first principles, and does not require any experimental input to make predictions about these materials.

We have used this new method to obtain information on LiNiPO$_4$ and on several silicate phosphates, for which no experimental information is available. We find that all these materials have high Li extraction potentials, with LiMnSiO$_4$ the lowest potential at about 4.8 V.   While they have very high extraction potentials, $LiFeSiO_4$ and $LiCoSiO_4$ are attractive due to their very low volume change with Li removal.   Li extraction from $LiFeSiO_4$ may be thermodynamically possible (4.9V) at the limit of current electrolytes.

*Corresponding Author: gceder@mit.edu
+1-617-253-1581 (phone) 617-258-6534 (fax)

Assuming these silicates can be synthesized in the olivine structures, a significant limitation may be their electronic conductivity.   In $LiFePO_4$ electron transport is a significant limitation to high power rate, which can only be overcome by particle coating methods [37, 38] or doping [39].   It is not yet clear whether similar methods are applicable to silicates.

## Acknowledgments

F. Z. wants to thank D. Morgan and T. Maxisch for helpful discussions of the olivine materials. This work was supported by the Department of Energy under contract numbers DE-FG02-96ER45571 and BATT program 6517748, and by the MRSEC program of the National Science Foundation under contract number DMR-0213282.

*Corresponding Author: gceder@mit.edu
+1-617-253-1581 (phone) 617-258-6534 (fax)

| | $U$ (eV) | | a (Å) | b (Å) | c (Å) | V (Å$^3$) | \|μ\| (μ$_B$) | Voltage (V) |
|---|---|---|---|---|---|---|---|---|
| LiMnPO$_4$ | 3.92 | GGA | 10.55 | 6.13 | 4.78 | 309.13 | 4.47 | 2.98 |
| | | GGA+$U$ | 10.62 | 6.17 | 4.80 | 314.52 | 4.65 | 4.04 |
| | | Exp.[32] | 10.44 | 6.09 | 4.75 | 302.0 | | 4.1 |
| MnPO$_4$ | 5.09 | GGA | 9.92 | 6.01 | 4.93 | 293.92 | 3.56 | |
| | | GGA+$U$ | 9.98 | 6.07 | 4.96 | 300.47 | 3.95 | |
| | | Exp.[32] | 9.69 | 5.93 | 4.78 | 274.7 | | |
| LiFePO$_4$ | 3.71 | GGA | 10.39 | 6.04 | 4.75 | 298.09 | 3.54 | 2.99 |
| | | GGA+$U$ | 10.42 | 6.07 | 4.76 | 301.07 | 3.73 | 3.47 |
| | | Exp.[10] | 10.33 | 6.01 | 4.69 | 291.4 | | 3.5 |
| FePO$_4$ | 4.90 | GGA | 9.99 | 5.93 | 4.90 | 290.28 | 3.93 | |
| | | GGA+$U$ | 9.99 | 5.88 | 4.87 | 286.07 | 4.33 | |
| | | Exp.[10] | 9.82 | 5.79 | 4.79 | 272.4 | | |
| LiCoPO$_4$ | 5.05 | GGA | 10.30 | 5.93 | 4.75 | 290.13 | 2.54 | 3.70 |
| | | GGA+$U$ | 10.33 | 5.97 | 4.76 | 293.55 | 2.78 | 4.73 |
| | | Exp.[40] | 10.20 | 5.92 | 4.70 | 283.9 | | 4.8 |
| CoPO$_4$ | 6.34 | GGA | 9.71 | 5.48 | 4.59 | 244.24 | 0 | |
| | | GGA+$U$ | 9.98 | 5.78 | 4.74 | 273.42 | 3.24 | |
| | | Exp.[40] | 10.09 | 5.85 | 4.72 | 278.7 | | |
| LiNiPO$_4$ | 5.26 | GGA | 10.09 | 5.91 | 4.74 | 282.66 | 1.54 | 4.20 |
| | | GGA+$U$ | 10.12 | 5.90 | 4.73 | 282.42 | 1.78 | 5.07 |
| | | Exp.[27] | 10.03 | 5.85 | 4.68 | 274.5 | | |
| NiPO$_4$ | 6.93 | GGA | 9.66 | 5.72 | 4.71 | 260.25 | 0.64 | |
| | | GGA+$U$ | 9.92 | 5.82 | 4.84 | 279.43 | 1.78 | |

Table 1: Calculated $U$ parameter, lattice parameters, unit cell volume, magnetic momentum on the transition metal ions, and voltage for olivine phosphates (Li)MPO$_4$.

*Corresponding Author: gceder@mit.edu
+1-617-253-1581 (phone) 617-258-6534 (fax)

|  | $U$ (eV) |  | a (Å) | b (Å) | c (Å) | V (Å$^3$) | \|μ\| (μ$_B$) | Voltage (V) |
|---|---|---|---|---|---|---|---|---|
| LiMnSiO$_4$ | 6.33 | GGA | 9.97 | 5.95 | 4.81 | 285.45 | 3.68 | 3.66 |
|  |  | GGA+$U$ | 10.03 | 6.02 | 4.82 | 291.03 | 4.08 | 4.87 |
| MnSiO$_4$ | 5.32 | GGA | 10.02 | 5.50 | 4.69 | 258.74 | 2.67 |  |
|  |  | GGA+$U$ | 10.09 | 5.72 | 4.81 | 277.61 | 3.75 |  |
| LiFeSiO$_4$ | 4.53 | GGA | 10.18 | 5.91 | 4.81 | 289.31 | 4.06 | 3.87 |
|  |  | GGA+$U$ | 10.17 | 5.87 | 4.80 | 286.55 | 4.35 | 4.97 |
| FeSiO$_4$ | 5.00 | GGA | 9.96 | 5.80 | 4.87 | 281.40 | 3.16 |  |
|  |  | GGA+$U$ | 10.13 | 5.76 | 4.90 | 286.14 | 4.09 |  |
| LiCoSiO$_4$ | 5.95 | GGA | 9.89 | 5.56 | 4.69 | 257.98 | 0 | 4.27 |
|  |  | GGA+$U$ | 10.14 | 5.78 | 4.76 | 279.15 | 3.27 | 5.10 |
| CoSiO$_4$ | 6.26 | GGA | 9.83 | 5.37 | 4.62 | 244.24 | 0.57 |  |
|  |  | GGA+$U$ | 9.93 | 5.75 | 4.90 | 279.78 | 3.16 |  |
| LiNiSiO$_4$ | 6.97 | GGA | 9.87 | 5.71 | 4.74 | 267.56 | 0.81 | 4.60 |
|  |  | GGA+$U$ | 10.07 | 5.80 | 4.80 | 280.40 | 1.64 | 5.18 |
| NiSiO$_4$ | 6.80 | GGA | 9.86 | 5.44 | 4.63 | 248.32 | 0 |  |
|  |  | GGA+$U$ | 10.15 | 5.55 | 4.79 | 269.83 | 1.56 |  |

Table 2: For olivine silicates (Li)MSiO$_4$.

*Corresponding Author: gceder@mit.edu
+1-617-253-1581 (phone) 617-258-6534 (fax)

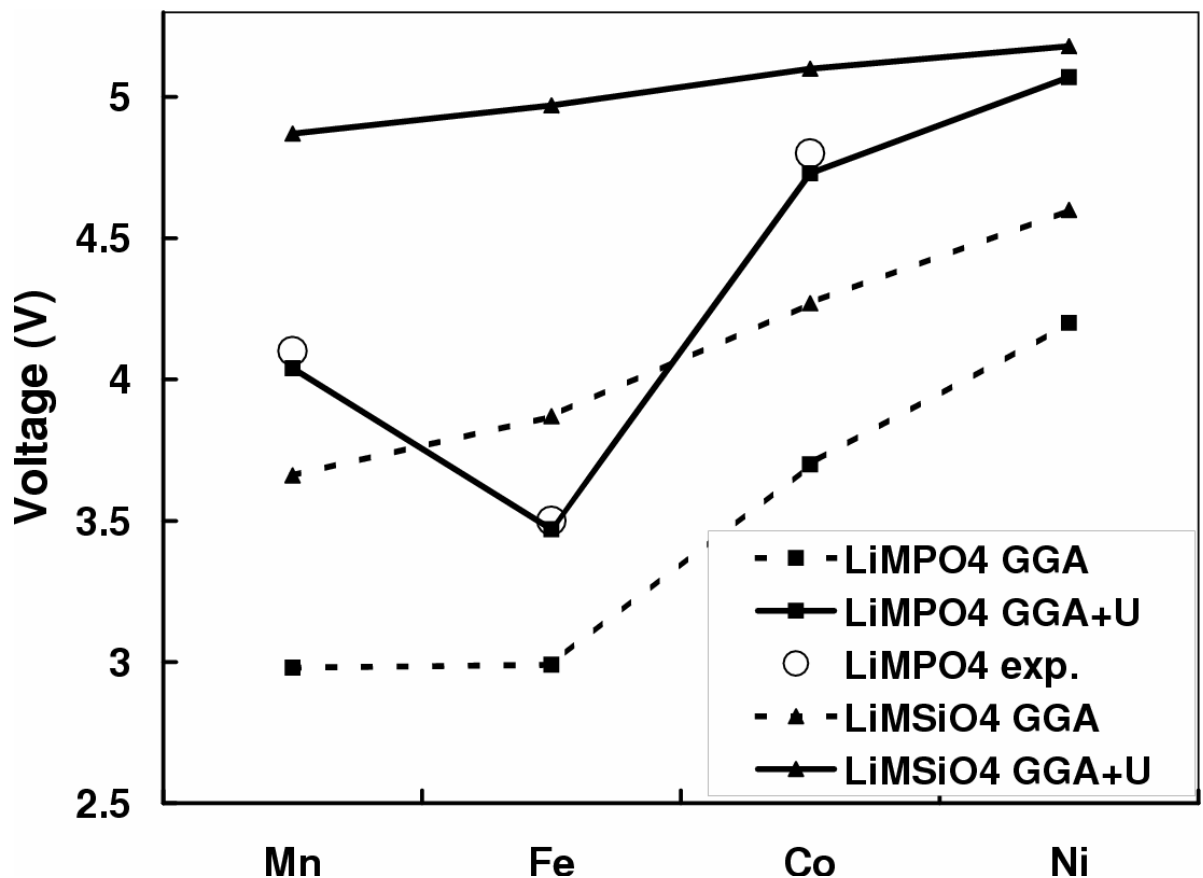

Figure 1. Voltage for M= Mn, Fe, Co and Ni in olivine phosphates LiMPO4 (squares) and silicates LiMSiO4 (triangles), as predicted by GGA (dotted line) and GGA+$U$ (solid line). Available experimental values for $LiMPO_4$ (M= Mn, Fe, Co) are also shown as big circles.

*Corresponding Author: gceder@mit.edu
+1-617-253-1581 (phone) 617-258-6534 (fax)

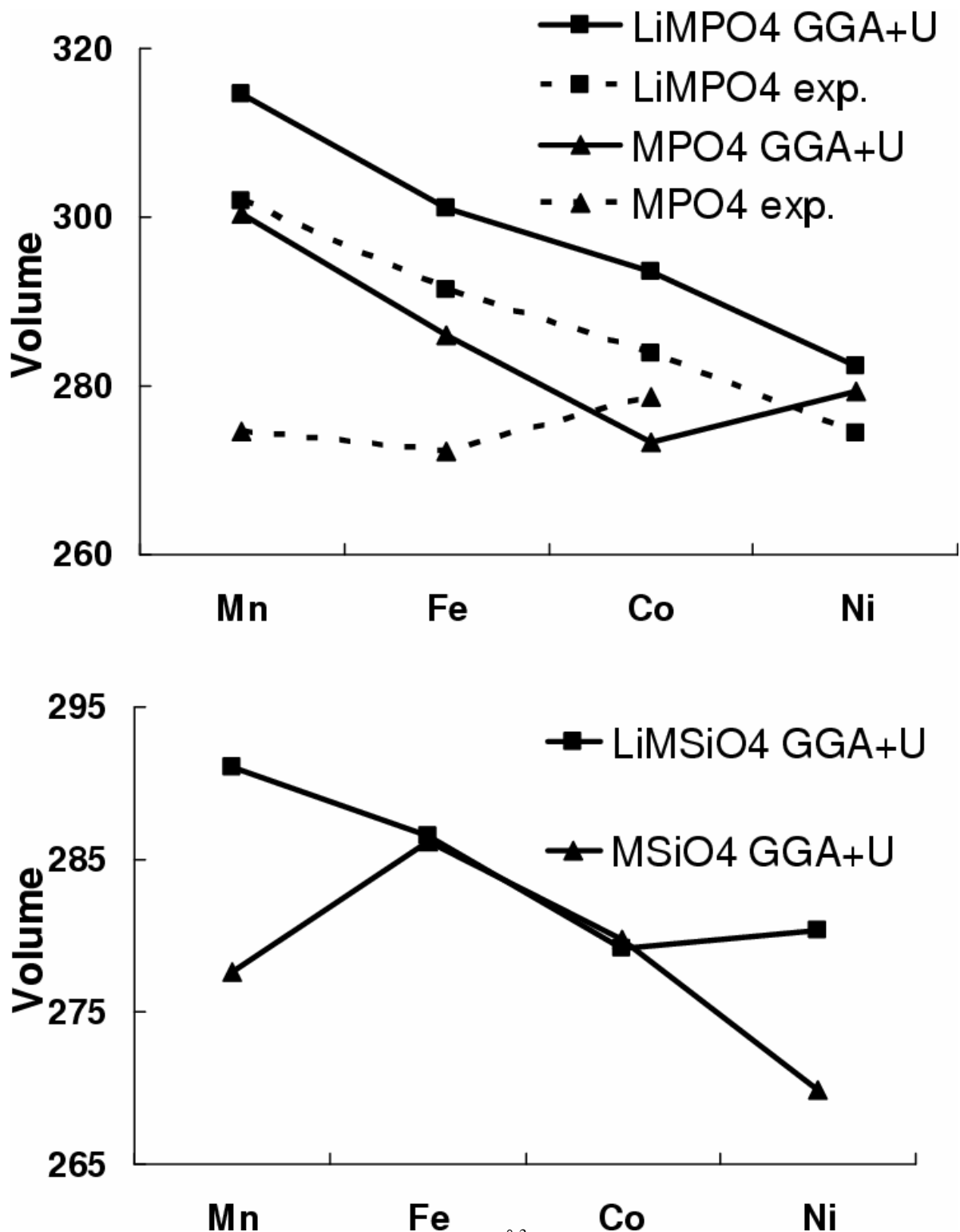

Figure 2. GGA+$U$ predicted unit cell volume in Å$^3$ for olivine phosphates and silicates, respectively. For phosphates experimental volume is also shown where available. For numbers see Table 1 and 2.

References

*Corresponding Author: gceder@mit.edu
+1-617-253-1581 (phone) 617-258-6534 (fax)

*Corresponding Author: gceder@mit.edu
+1-617-253-1581 (phone) 617-258-6534 (fax)

*Corresponding Author: gceder@mit.edu
+1-617-253-1581 (phone) 617-258-6534 (fax)